\documentstyle[11pt,psfig]{article}
\input{epsf}
\begin{document}
\pagenumbering{arabic}
\begin{center}
\title{GENERAL STATISTICAL PROPERTIES OF THE CMB POLARIZATION FIELD}
\author{\hspace{0.5cm} P.D. Naselsky\\
Rostov State University, Zorge 5, Rostov-Don, 344104 Russia,\\
Theoretical Astrophysics Center, Juliane Maries Vej 30, 2100 Copenhagen,
Denmark.\\
and\\
D.I.Novikov\\
University of Kansas, Dept. of Physics and Astronomy,
Lawrence, Kansas, 66045\\
University Observatory, Juliane Maries Vej 30, 2100 Copenhagen, Denmark.}
\date{}
\maketitle
\end{center}
\begin{abstract}

The distribution of the polarization of the Cosmic Microwave
Background (CMB) in the sky is determined
by the hypothesis of random Gaussian distribution of the primordial
density perturbations. This hypotheses is well motivated by the
inflationary cosmology. Therefore, the test of
consistency of the statistical properties of the CMB polarization
field with the Gaussianity of primordial density fluctuations
is a realistic way to study the nature of primordial
inhomogeneities in the Universe.

This paper contains the theoretical predictions of the general
statistical properties of the CMB polarization field. All results
obtained under assumption of the Gaussian nature of the signal.
We pay the special attention to the following two problems. First, the
classification and statistics of the singular points
of the polarization field where polarization is equal to zero. Second, the 
topology of contours of the value of the degree of polarization. We have
investigated the percolation properties for the zones of ``strong''
and ``weak'' polarization. We also have calculated Minkowski
functionals for the CMB polarization field. All results are analytical.

\vspace{0.3cm}

{\it Subject headings:} cosmic microwave background, cosmology, statistics,
observations.
\end{abstract}

\section{Introduction}

Observations of the anisotropy and polarization of the  Cosmic Microwave
Background (CMB) provide a unique information about the primordial 
inhomogeneity of the Universe. Since detection by COBE 
(Smoot et al, 1992, Bennett et al, 1996) of the 
CMB anisotropy several groups  have reported on high angular 
resolution observational data at the angular  scales $\theta\sim1^o$ in the 
vicinity of the so called Doppler peak in the $\Delta T/T$ power spectrum
(Hancock, et al, 1994;
Gundersen, 1993;
De Bernarais, et al., 1994;
Masi, et al, 1996;
Tanaka, et al., 1995;
Cheng, et al., 1994;
Netterfield  et al., 1996;
Scott, et al, 1996). Determination of the spectrum 
of the primordial anisotropy on scales $\theta\sim 1^o$  will yield 
valuable clues to the formation of the large scale structure of the 
Universe and the most important parameters of the Universe: the total and 
baryonic densities at the present time  ($\Omega$ and $\Omega_b$); the  Hubble 
constant, ionization history etc.
However, the interpretation of these experimental results
as well as the comparison  with the expected power spectra of anisotropy 
in different cosmological models is complicated. Future experiments
(MAP and Planck) will construct the map of the CMB with high level of
resolution and sensitivity. 

The temperature distribution in the sky also contains the information
beyond power spectrum. Gaussianity of the primordial signal
is well motivated by the inflationary cosmology and has been adopted
by many authors (see for review Starobinsky 1982, Bardeen et al. 1983, 
Bardeen et al. 1986). In this case the distribution of the CMB temperature
in the sky is in the form of the two-dimensional scalar random Gaussian field.
This field can be completely characterized by its power spectrum.
Many authors proposed to calculate various statistical characteristics
of the CMB anisotropy as tests of Gaussianity. All these techniques
provide information beyond power spectrum:

1. Statistic of peaks in a random Gaussian fields. Following
classical papers of Doroshkevich 1970 and 
Bardeen,J.M.,Bond,J.R.,Kaiser,N.,$\&$ Szalay, 1986 - (BBKS)),
this
approach has been developed by (Bond and Efstathiou 1987,
P.Coles 1988) for CMB anisotropy.

2. Higher-order correlations - 3, 4 ets.(Luo and Schramm 1994,
Smoot et al. 1994, Kogut et al. 1996);

3. Minkowski
functionals as a morphological descriptors of the CMB anisotropy
maps (Shmalzing and Gorski 1997, Winitzki and Kosowsky 1997).
As it was mentioned by Shmalzing and Gorski 1997, Minkowski 
functionals are very sensitive to non-Gaussianity. This
approach is very effective because these functionals are
additive with respect to the isolated regions in the sky and they
have simple analytical form in the case of the
Gaussian field. This approach can be used to test predictions
of Gaussianity. 

4. Percolation and cluster analysis.
This statistical method based on a very attractive idea:
if the experimental signal is a sum
of the primordial signal and non-Gaussian noise (for example foreground
sources, dust emission and so on), then statistical properties of the pure 
Gaussian signal could be distorted. This effect provides a basis for the 
investigation of the characteristics of the non-Gaussian noise in the 
experimental data. Note, that percolation has become a popular
term among cosmologists. The percolation technique has been 
successfully applied for investigation of the evolution of
the spatial density
distribution in the Universe due to gravitational instability 
(see for review Zeldovich 1982, Shandarin 1983,
Dominik, $\&$ Shandarin 1992). For CMB anisotropy this technique
has been developed by (Naselsky and Novikov D. 1995,
Novikov D. and Jorgensen 1996).  

Therefore, the data analysis of the CMB anisotropy can be
divided into two parts: power spectrum estimation with subsequent
cosmological parameters extraction, and investigation of the nature of the
observed signal. 

There is another important characteristic of the distribution
of the CMB on the sky: this is the CMB polarization.

The idea that polarization provides important information about the 
primordial cosmic plasma was pointed out by Rees (1968). The 
properties of the power spectrum of the CMB polarization field were
analyzed in for example
(Basco and Polnarev 1979, Polnarev 1985, Bond and Efstathiou 1987,
Coulson et al. 1994, Crittenden et al. 1995,
Zaldarriaga and Harari 1995, Ng K.L. and Ng K.W., 1995, Kosowsky 1996;
Kamionkowski et al. 1996, Jungman et al. 1996,
Naselsky and Polnarev 1987, Ng K.L. and Ng K.W. 1996, Hu and White, 1997). 

The polarization field also contains information beyond power spectrum
which also can be used for
investigation of the nature of the primordial inhomogeneity in the
Universe. Statistical properties of the polarization field caused by
Gaussian fluctuations was partly
discussed by Bond and Efstathiou 1987, Arbuzov et al. 1997a, 
Arbuzov et al. 1997b. It is important to note, that polarization
contains more information about nature of the primordial signal than
the anisotropy (the polarization field is a combination of two random
independent Gaussian fields (Bond and Efstathiou 1987, while anisotropy
of the CMB is only one). 

In this paper we focus attention on the general statistical properties
of the CMB polarization field. This is not a scalar field
(unlike the anisotropy) and can be completely described in terms of
Stokes parameters - $Q$, $U$ and $V$. Since Thomson scattering does not
produce circular polarization, $V=0$ we can consider the level of
polarization, which depends only on two parameters -
$\Pi=\frac{\sqrt{Q^2+U^2}}{I}$, where I is the total intensity, and
$P=\Pi \times I=\sqrt{Q^2+U^2}$ is the polarized intensity. Therefore,
polarization field can be described in terms of the angle of
polarization $2\varphi=$arctg$\frac{Q}{U}$ and polarized intensity -
P.  Since polarization of the radiation does not have any direction
(it has only the orientation $\varphi$ and intensity P), it cannot be
formally interpreted as a vector field. Nevertheless, below we use the term
``vector of polarization'' $\overline{P}$ (so that 
$P=|\overline{P}|$) for simplicity, taking into
account that this ``vector'' is not directed.  We assume that $Q$ and
$U$ components of the autocorrelated pseudo-vector $\overline{P}$ are
statistically independent (Bond and Efstathiou 1987) and have a
Gaussian distribution on the
sky. We are interested in the general
statistical properties of the $\overline{P}$
distribution such as surface density and classification
of the non-polarized points $P=0$ in the sky, Minkowski functionals
for the value of $P$ and the percolation of the relatively 
strongly polarized spots.

\section{Pattern of the polarization fluctuations}

In this section we discuss very specific features of the
polarization pattern of the cosmic microwave background. All results
were obtained under the assumption that the polarization field is the
result of a random Gaussian process. We describe the statistical
properties of the two dimensional vector field of the polarization
$\overline{P}$ such as the surface density of the singular points 
$\overline{P}=0$
(section 2.1),
genus curve for the two dimensional 
scalar field $|\overline{P}|$ and the level of percolation through the 
relatively strongly polarized 
spots (section 2.2).
In this section we consider small angular parts of the sky without
loss of generality. Thus 
the geometry is approximately flat and the vector of polarization can be
described in the following form:

\begin{equation}
\overline{P}=P_x\overline{x}+P_y\overline{y},
\end{equation}

where $\overline{x}$ and $\overline{y}$ are the unit
vectors of the Cartesian coordinate
system on the small angular part of the unit sphere and components $P_x$
and $P_y$ can be expressed in terms of Stokes parameters $Q$ and $U$:

\begin{equation}
\begin{array}{l}
P_x=Pcos(\varphi )\\
P_y=Psin(\varphi )\\ 
Q=Pcos(2\varphi )\\
U=Psin(2\varphi )
\end{array}
\end{equation}

Where $\varphi$ is the orientation of polarization and:

\begin{equation}
\begin{array}{l} 
P_x=cos(\varphi )Q-sin(\varphi )U\\
P_y=sin(\varphi )Q+cos(\varphi )U
\end{array}
\end{equation}

Therefore, components $P_x$ and $P_y$ are also independent random 
two-dimensional Gaussian fields with the same parameters as $Q$ and $U$.
It means, that the statistical properties of the vector 
$Q\overline{x}+U\overline{y}$ are equivalent to the properties
of the vector $\overline{P}$. It allows us to use the usual terms
$Q$ and $U$ instead of $P_x$ and $P_y$. 

\subsection{Singular points of the polarization vector field.}

First, we are interested in the statistics of the singular points $(x_0,y_0)$
of the vector $\overline{P}$: $\overline{P}(x_0,y_0)=0$. 
This condition means, that both 
components $Q$ and $U$ are equal to zero in such points simultaneously:
\begin{equation}
Q(x_0,y_0)=U(x_0,y_0)=0.
\end{equation}
The surface density of these points can easily be computed 
analytically. Points $(x_0,y_0)$ are the points of the intersection of the 
lines of zero level of the $Q$ and $U$ surfaces. The angular density of such 
points can be found by using the properties of the joint probability function 
for  distribution of $Q$, $U$, $Q_1$, $U_1$, $Q_2$, $U_2$.
Here $Q_i$ and $U_i$ are the first derivatives of $Q$ and $U$ respectively
in the point $(x_0,y_0)$:
\begin{equation}
f_1(x_0,y_0)=\left.\frac{\partial f}{\partial x}\right|_{x_0,y_0} \hspace{1cm}
f_2(x_0,y_0)=\left.\frac{\partial f}{\partial y}\right|_{x_0,y_0} .
\end{equation}
These 6 different values are independent (Bardeen et al. 1986) for an 
arbitrary point $(x,y)$ of the map  and have zero average and 
the following variances:
\begin{equation}
\langle Q^2\rangle=\langle U^2\rangle=\sigma^2_0, \hspace{0.5cm} 
\langle Q^2_i\rangle=\langle U^2_i\rangle=\sigma^2_1/2,\hspace{0.5cm} i=1,2.
\end{equation}
where $\sigma_o$ and $\sigma_1$ are the spectral parameters, as they
were defined by Bond and Efstaphiou 1987.
The  joint probability for these values is:
\[
X(Q,U,Q_i,U_i)dQdUdQ_idU_i=\frac{4}{\sqrt{(2\pi)^6}
\sigma_0\sigma_1^2}e^{-\frac{1}{2}A}dQdUdQ_idU_i
\]
\begin{equation}
A=\frac{Q^2}{\sigma_0^2}+\frac{U^2}{\sigma_0^2} +
2\sum_{i=1,2}\left(\frac{Q^2_i}{\sigma_1^2}+\frac{U^2_i}
{\sigma_1^2}\right).
\end{equation}
In the vicinity of the singular point ($x_0,y_0$), the value of 
$Q$ and $U$ can be described by the following expression:
\begin{equation}
\left(
\begin{array}{c}
Q \\ U 
\end{array}
\right)=\left(
\begin{array}{cc}
Q_1 & Q_2\\
U_1 & U_2\\
\end{array}
\right)\times\left(
\begin{array}{c}
\Delta x \\ \Delta y
\end{array}
\right)
\end{equation}
The substitution  of $dQdU=|\det(Q_i,U_i)|dxdy$ and integration 
over $dQ_idU_i$ gives us the number density of the singular points:
\begin{equation}
N_{np}=\frac{4}{(2\pi)^3}\frac{\sigma_1^2}{\sigma_0^2}\int e^{-(q_i^2+u_i^2)}
|q_1u_2-u_2q_1|dq_idu_i
\end{equation}
 where $q_i=\frac{Q_i}{\sigma_1}$, $u_i=\frac{U_i}{\sigma_1}$, $i=1,2$.

The analytical calculation of the surfase density of the singular points 
can be found in Appendix A. Here we present  the main results and 
conclusions only.

\noindent
{\bf a. Classification of singular points}

We investigate the polarization vector field around the singular points 
in the following way. Let us imagine that a point in the vicinity of the 
singular point is moving along the lines of the vector field. In this 
case  the investigation is similar to that  for the 
singular points of linear differential equations. Following Eq.(5)
the field in the small vicinity of the point $(x_0,y_0)$, where 
$q(x_0,y_0)=u(x_0,y_0)=0$ can be described in terms of the matrix $M$ of 
first derivatives of the field $q$ and $u$, and we can consider the following 
equation
\begin{equation}
\left(
\begin{array}{c}
\dot{x} \\  \dot{y}
\end{array}
\right)
= M
\left(
\begin{array}{c}
x \\  y
\end{array}
\right),
\end{equation}
where $x=x_0+\Delta x$, $y=y_0+\Delta y$, and
\[
M=\left(
\begin{array}{cc}
q_1 & q_2\\
u_1 &  u_2
\end{array}
\right).
\]
The characteristic equation for Eq.(10) is:
\begin{equation}
\lambda^2-(q_1+u_2)\lambda +(q_1u_2-q_2u_1)=0.
\end{equation}
This equation has two roots: $\lambda_1$ and $\lambda_2$, ordered by
$Re(\lambda_1 )> Re(\lambda_2 )$,
and the classification 
of the singular point $(x_0,y_0)$ depends on their values.\\
1. If $Im(\lambda_1)=-Im(\lambda_2)\neq 0$, then this is a
focus and the vector 
field will spiral toward the point $(x_0,y_0)$ (Fig.1 left).\\
2. If $\lambda_1$ and $\lambda_2$ are real, then the matrix
$M$ has two eigenvectors which correspond to different values
$\lambda_1$ and $\lambda_2$, and we 
can consider two different cases:

\hspace{1cm} {\bf a)}. $\lambda_2>0$ - both values are positive 
or $\lambda_1<0$ -
both values are negative. This means, that the point $(x_0,y_0)$ is a knot 
and the lines of the vector field tend to be aligned to the direction
of the eigenvector with maximal value of
$|\lambda_i|$, $i=1,2$ (Fig.1 middle).

\hspace{1cm} {\bf b)}. $\lambda_1>0$ and $\lambda_2<0$ -
values with opposite signs. 
In this case the point $(x_0,y_0)$ is a saddle (Fig.1 right).

Singular points of different types determine the behavior of the vector field
in their vicinities. The distribution of the singular points on the map of
cosmic microwave background polarization determines the topology of the 
relatively small polarized zones Fig.2.

\noindent
{\bf b. Surface density of the singular points.}

Detailed calculation of the surfase density of the singular points 
is in the Appendix A. 
The surface density of singular points of different kinds are:
\begin{equation}
\begin{array}{l}
N_f=\frac{\sqrt{2}}{16\pi}\frac{1}{r_c^2},\\
N_k=\frac{\sqrt{2}}{16\pi}\left(\sqrt{2}-1\right)\frac{1}{r_c^2},\\
N_s=\frac{1}{8\pi r^2_c},\\
\end{array}
\end{equation}
where $N_f$, $N_k$, $N_s$ are the number densities of 
focuses, knots and saddles
respectively and $r_c=\frac{\sigma_0}{\sigma_1}$ is the correlation radius.
The total number density of non-polarized points is:
\begin{equation}
N_{np}=N_f+N_k+N_s=\frac{1}{4\pi r_c^2}.
\end{equation}
Density of the singular points depends essentially on
correlation radius (Eq. 12, 13) and therefore on the spectral parameters 
$\sigma_0$, $\sigma_1$. These spectral parameters depend on the spectra of
polarization and on the device resolution (Bond and Efstathiou, 1987)
(see also Fig.3). The ratios
\begin{equation}
N_f/N_k=\sqrt{2}+1 \hspace{1cm} and \hspace{1cm} N_f/N_s=\frac{\sqrt{2}}{2}
\end{equation}
are the spectral independent constants determined only by the Gaussian 
nature of the primordial inhomogeneity in the Universe.
These ratios are a characteristic feature of the CMB polarization
vector field in the inflationary cosmology. Note, that for example
in the two-dimensional potential vector field 
($\overline{V}=\bigtriangledown \alpha$)
the number density of foci is equal to zero since this field does not
have a rotational component. 

The joint probability for the distribution of the eigenvalues of 
$\lambda_1$ and 
$\lambda_2$ in the singular points which can be either knots or
saddles (not foci) is:
\begin{equation}
F(\lambda_1,\lambda_2)d\lambda_1d\lambda_2=
\frac{8}{4-\sqrt{2}}\frac{1}{\sqrt{\pi}}
|\lambda_1\lambda_2|(\lambda_1-\lambda_2)
e^{-\frac{\lambda_1^2}{2}-\frac{\lambda_2^2}{2}}d\lambda_1d\lambda_2.
\end{equation}
Note, that this distribution is universal for all kinds of spectra of 
polarization as well as ratios (15).

\subsection{Percolation pattern for polarization}

Here we present our results for the Genus statistics of the value
$p=\frac{|\overline{P}|}{\sigma_0}$.
We can divide the map of polarization of the CMB into two parts: regions with
relatively strong polarization $p>p_0$ 
(``strongly polarized zones'') and 
regions with relatively weak polarization $p<p_0$ 
(``weakly polarized zones''). Below we find the value $p_0$ where percolation
through the ``strongly polarized zones'' changes to percolation through the 
``weakly polarized zones'' (Fig. 3,4). Let us suppose that we can
measure only a signal with polarized intensity $p\geq p_t$, where $p_t$ is
the threshold which
determines by the sensitivity of the device.
If we can measure only "strongly polarized" 
signal - $p_t>p_0$, then we can see only the separated polarized spots
which do not percolate. 
Therefore, the
percolation trough polarized zones can be reached only by the device
with the sensitivity $p_t\leq p_0$.

The value $p_0$ can be found analytically in the following way. We consider 
the value $|\overline{P}|$ as a two-dimensional random scalar
field with a Rayleigh distribution
(Coles and Barrow, 1987). This field can be imagined as a two-dimensional
surface in a three-dimensional space. This surface has extreme points such as 
maxima, minima, saddle points and singular points. The last ones have been
considered in the previous subsection. The densities of maxima, minima and
saddle points have some distributions with $p$:
\begin{equation}
\begin{array}{l}
N_{max}(p)=\int\limits_{p}^{\infty}n_{max}(p')dp',\\
N_{min}(p)=\int\limits_{p}^{\infty}n_{min}(p')dp',\\
N_{sad}(p)=\int\limits_{p}^{\infty}n_{sad}(p')dp',
\end{array}
\end{equation}
where $n_{max}(p)$, $n_{min}(p)$, $n_{sad}(p)$ - are the number densities of
maxima, minima and saddle points respectively on some interval - (p,p+dp),
and $N_{max}(p)$, $N_{min}(p)$, $N_{sad}(p)$ are
the number densities of maxima, minima and saddle points 
respectively above some level $p$. (We note, that here saddle points are
the saddle points of the two-dimensional surfase of p(x,y). These
points are not the same as saddle (kind of singular points)
in the previous section.)\\
The definition of the Genus  is:
\begin{equation}
g(p)=n_{max}(p)+n_{min}(p)-n_{sad}(p).
\end{equation}
The integrated Genus is then:
\begin{equation}
G(p)=N_{max}(p)+N_{min}(p)-N_{sad}(p)=
\int\limits_{p}^{\infty}g(p')dp'.
\end{equation}
The level of percolation  $p_0$ has to be found from the condition
$G(p_0)=0$. We recognize that this condition does not automatically
mean that $p_0$ is the level of percolation for an arbitrary scalar
field. It is well-known that for the Gaussian random field the percolation
level corresponds to the level where Genus curve intersects the zero.
We have checked this condition for the Rayleigh distribution
by simulating a large number of realizations for a
two-dimensional field. In the case of Rayleigh   distribution this
condition also mean that level $p_0$ corresponds to the percolation contour.

The detailed calculation of the Genus can be found in Appendix B. 
Formally the steps of its calculation are as follows:\\
1. The value $p$ is a combination of the independent random values
$q$ and $u$. 
The first and second derivatives of them are: $q_i$, $u_i$, $q_{ij}$, 
$u_{ij}$, ($q_{ij}=Q_{i,j}/\sigma_{2},u_{ij}=U_{ij}/\sigma_{2}$),
$i=1,2$, where $\sigma_2$ is also the spectral parameter as it was
defined by Bond and Efstaphiou 1987: $\sigma_2^2=\langle Q_{ii}^2\rangle =
\langle U_{ii}^2\rangle$. These values obey the following conditions:
\begin{equation}
\begin{array}{l}
p^2=q^2+u^2\\
p_i=qq_i+uu_i,\\
\gamma p_ip_j+p_{ij}=\gamma (q_iq_j+u_iu_j)+qq_{ij}+uu_{ij},\\
\langle qu\rangle =\langle q_iu_j\rangle=\langle q_{ij}u_{kl}\rangle=
\langle qu_{i}\rangle=\langle q_{i}u\rangle=0,\\
\langle q q_{ij}\rangle = \langle u u_{ij}\rangle=-\frac{\gamma}{2}
\delta_{ij},\\
\langle q_iq_j\rangle=\langle u_iu_j\rangle=\frac{1}{2}\delta_{ij},\\
\langle q_{ij}u_{kl}\rangle=
\frac{1}{8}(\delta_{ik}\delta_{jl}+\delta_{il}\delta_{jk}+\delta_{ij}
\delta_{kl}),\\
\gamma=\frac{\sigma_1^2}{\sigma_{0} \sigma_{2}}
\end{array}
\end{equation}

\noindent
2. The joint probability F of the Gaussian distribution for the values
$q$, $q_i$, $q_{ij}$,  $u$, $u_{i}$, $u_{ij}$ is:
\begin{equation}
\begin{array}{l}
Fdqdudq_idu_idq_{ij}du_{ij}=\frac{1}
{\sqrt{(2\pi)^{12}\det M}}e^{-\frac{A}{2}}dqdudq_idu_i
dq_{ij}du_{ij},\\
A=v\times M^{-1}\times v^T
\end{array}
\end{equation}
where $M$ is the covariance matrix and
A is the quadratic form of the 
12-dimensional vector $v=(q,q_i,q_{ij},u,u_{i},u_{ij})$.

\noindent
3. The substitution of $p$, $p_i$, $p_{ij}$ in Eq.(20) from Eq.(19) 
and integration over 6 variables gives us the joint probability 
$fdpdp_idp_{ij}$ for values $p$, $p_i$, $p_{ij}$ to be in the range from 
$p$, $p_i$, $p_{ij}$ to $p+dp$, $p_i+dp_i$, $p_{ij}+dp_{ij}$.

\noindent
4. The differential density of the extreme points obeys the equation:
\begin{equation}
n_{ext}(p)=\frac{\sigma_2^2}{\sigma_1^2} \int |
\det(p_{ij})|f\delta(p_1)\delta(p_2)dp_{ij},
\end{equation}
where $n_{ext}(p)$ is the density of the extreme points.
These extreme points can be maxima, minima or saddle points
depending on the limits of the integration over $du_{ij}$.
These limits determine by the values of $tr(p_{ij})$ and $det(p_{ij})$
of the second derivatives matrix $(p_{ij})$ (see Appendix B).

\noindent
5. The Genus curve obeys the equation:
\begin{equation}
g(p)=n_{max}(p)+n_{min}(p)-n_{sad}(p)=\frac{\sigma_2^2}{\sigma_1^2}
\int\det (p_{ij}) f(p,p_i=0,p_{ij})dp_{ij}.
\end{equation}
After integrating this, we have
\begin{equation}
g(p)=\frac{1}{4\pi}\left(\frac{\sigma_1}{\sigma_0}\right)^2p(p^2-3)
e^{-\frac{p^2}{2}}.
\end{equation}
The integrated Genus curve is
\begin{equation}
G(p)=\frac{1}{4\pi r_c^2}(p^2-1)e^{-\frac{p^2}{2}},
\end{equation}
Condition $G(p)=0$ gives us the value of $p_0$:
\begin{equation}
p_0=1.
\end{equation}
Taking into account that random value $p$ has distribution
$pe^{-\frac{p^2}{2}}dp$ we 
can obtain that percolation through the ``strongly polarized'' zone when 
a part $e^{-\frac{p^2_0}{2}}$ of the map is detected as a ``strongly
polarized''.  This 
corresponds to $\approx 61\%$ of the map's area.

When $p_0=0$ in Eq.(25) we have
\begin{equation}
g(0)=-\frac{1}{4\pi r_c^2}.
\end{equation}
This value exactly coincides with $N_{np}$ in Eq. (13) with the
opposite sign. 
These null-points are the non-smooth minima of the surface $p$. 
The non-smooth minima have not been taken into account
in Eqs.(16)-(25). Therefore, the total number of minima per
unit area is $N_{min total}=N_{np}+N{min}(0)$, where $N_{np}$ are minima,
if p=0 and $N_{min}(0)$ are minima, if p>0. Therefore the total number of
extreme points per unit area are:
\begin{equation}
\begin{array}{l}
N_{max total}=N_{max}(0),\\
N_{sad total}=N_{sad}(0),\\
N_{min total}=N_{np}+N{min}(0).
\end{array}
\end{equation}  
Taking into account equations (13,18,25,27) we obtain:
\begin{equation}
N_{max total}+N_{min total}-N_{sad total}=0,
\end{equation}
as it should be.

\section{Minkowski functionals for CMB polarization field}

As it was mentioned above, the CMB polarization at any point
of the map can be
characterized by the orientation angle and polarized intensity - p.
This intensity has the random Rayleigh distribution on the 
sky - $p=\sqrt{q^2+u^2}$. Therefore, the value of $p$ can be considered
as a two-dimensional random Raleigh field. It is well-known, that
two-dimensional field has only three Minkowski functionals which satisfy 
additivity and translational invariance (Minkowski 1903, Hadwiger 1959).

Geometrical interpretation of the Minkowsky functionals on the 
two-dimensional map is essentially easy. Analogously to the
previous section, we consider polarized intensity as a two-dimensional
surface in a three-dimensional space. If we cut this surface at the
different levels $p_t$, then the area of the map will be divided into
two parts: the area, where polarization is above the threshold $p_t$ and
the area, where $p<p_t$. 
For a two-dimensional distribution, Minkowski functionals correspond
to the following values:\\
1. A - fraction of the area of the map, where $p>p_t$;\\
2. L - length of the boundary between fractions, where $p>p_t$ and $p<p_t$
per unit area;\\
3. $G=N_{max}+N_{min}-N_{sad}$ - Euler characteristic 
(equivalent to the genus) per unit area.

Therefore, threshold is the independent variable on which these functionals
depend.
The third functional has already been considered in the previous section.
The obvious first one is $e^{-\frac{p_t^2}{2}}$. The second one can be
obtained in the same way as it was done for the the Gaussian field
(Adler 1981). Here we present the result without derivation:

\begin{equation}
L=\frac{1}{r_c}p_t e^{-\frac{p_t^2}{2}}
\end{equation}
  
The comparison of the Minkowski functionals for the CMB polarization
field with these functionals for CMB anisotropy is in the Fig. 5.
Functionals for Rayleigh distribution are equal to zero for $p_t<0$.
The third functional should be described together with the number of
non-polarized points (see previous section). 
These functionals can be used as a morphological descriptor of the
CMB polarization field in a similar manner as for
the CMB anizotropy (Winitski and Kosowski 1997).

\section{Discussions}

In this paper we have presented calculations of the statistical properties 
CMB polarization maps.

We believe that these statistical 
properties can be useful for checking the polarization patterns for 
presence of the non-Gaussian noise (for example, confusion signal from 
sources which can have the same spectral parameters as the polarization 
of the CMB). If an observational signal is free from non-Gaussian noise 
and is Gaussian itself (due to inflation) then the topological approach
is not necessary, 
because the correlation function or equivalent its power spectrum 
contains all information about the polarization signal. On the other hand, 
if the signal is a sum of polarization of the CMB (which is Gaussian) 
and unresolved foreground sources (which are non-Gaussian), then a detailed 
topological picture of the polarization field around the non-polarization 
points will be distorted comparative the predictions of the theory for the 
Gaussian distributions. 

On the other hand, the investigation of the nature of the primordial polarized
signal is a test of the inflationary model of the evolution of the Universe.
Therefore, the investigation of the Minkowski functionals together with
the non-polarized points on the observational data and comparison with
the theoretical predictions can be used as the test on Gaussianity of
the primordial inhomogeneity.

We would like to emphasize that the regions with strong polarization
will be detected easier than the regions with weak polarization. As we
demonstrated in the paper (see section 2.2) these region occupy an essential 
part of a whole map. From this point of view it is interesting to study the 
statistical properties of these regions. It is worth also mentioning that  
it is interesting to investigate the dependence of the spectral 
parameters of polarization in various cosmological models on the resolution 
of the detector and related statistical properties of the maps of 
polarization of the CMB. It is also very interesting to stady
the cross-correlations between anisotropy and polarization on the sky map
and make some theoretical predictions from the geometrical point of view.
These quations will be considered in a separate paper.

\vspace{1cm}

We would like to thank A. Melott, I. Novikov and S.Shandarin for stimulating 
discussion. At the University 
of Kansas the research was supported by NSF-NATO fellowship (DGE-9710914)
and by the NSF EPSCOR program.
We are grateful to the staff of TAC, and the Copenhagen 
University Observatory for providing excellent working conditions. This 
investigation was supported in part by the Russian Foundation for Fundamental 
Research (Code 96-02-17150) and by a grant ISF MEZ 300 as well as by the 
Danish Natural Science Research Council through grant No. 9401635 and also in
part by Danmarks Grundforskningsfond through its support for the 
establishment of the Theoretical Astrophysics Center.

\vspace{0.5cm}

\begin{center}
{\large\bf Appendix A}
\end{center}

\noindent
{\bf Surface density of the singular points}

\vspace{0.5cm}
\noindent
Below we describe the density of foci, knots and saddles. The total density of 
singular (non-polarized) points is:
\[ 
N_{np}=\frac{4}{(2\pi)^3}\left(\frac{\sigma_1}{\sigma_0}\right)^2
\int|q_1u_2-u_1q_2|e^{-\frac{A}{2}}dq_idu_i, \hspace{4.3cm}(A1)
\]
where
\[
A=\sum\limits_{i=1,2} 2q_i^2+2u_i^2, \hspace{0.5cm} i=1,2.
\]
The substitution 
\[
\begin{array}{ll}
q_1=\frac{1}{2}(x+R\cos\varphi), & u_1=\frac{1}{2}(w+R\sin\varphi),\\
q_2=\frac{1}{2}(R\sin\varphi-w), & u_2=\frac{1}{2}(x-R\cos\varphi),\\
\end{array}
\hspace{4.8cm}(A2)
\]
and integration over $d\varphi$ gives us:
\[
N_{np}=\frac{1}{4}\frac{1}{(2\pi)^2}\left(\frac{\sigma_1}{\sigma_0}\right)^2
\int e^{-\frac{1}{2}(x^2+R^2+w^2)}|x^2+w^2-R^2|R\ dRdxdw .\hspace{1cm}(A3)
\]
The next substitution $b=R^2-w^2$ allows us to rewrite (A3) in the following form:
\[
N_{np}=\frac{1}{8}\frac{1}{(2\pi)^2}\left(\frac{\sigma_1}{\sigma_0}\right)^2
\int e^{-\frac{1}{2}x^2}dx\int e^{-w^2}dw\int|x^2-b|e^{-\frac{b}{2}}db
\hspace{2cm}(A4)
\]
where $-w^2<b<\infty$, $-\infty<x<+\infty$, $-\infty<w<+\infty$.
In terms of $x$ and $b$ the eigenvalues of the matrix 
\[
M=\left(
\begin{array}{ll}
q_1 & q_2\\
u_1 & u_2
\end{array}\right)
\hspace{9.5cm}(A5)
\]
are
\[
\lambda_{1,2}=\frac{1}{2}(x\pm\sqrt{b}).\hspace{9.7cm}(A6)
\]
From (A4, A6) we can obtain the density of focuses, saddles and knots:
\[
\begin{array}{lll}
-w^2<b<0 & - & foci\\
0<b<x^2 & - & knots\\
x^2<b<\infty & - & saddles
\end{array}
\hspace{7.8cm}(A7)
\]
Using (A4), (A7) we obtain 
\[
\begin{array}{l}
N_f=\frac{\sqrt{2}}{16\pi}\left(\frac{\sigma_1}{\sigma_0}\right)^2\\
N_k=\frac{1}{16\pi}(2-\sqrt{2})\left(\frac{\sigma_1}{\sigma_0}\right)^2\\
N_s=\frac{1}{8\pi}\left(\frac{\sigma_1}{\sigma_0}\right)^2\\
\end{array}
\hspace{8.2cm}(A8)
\]
Using (A4, A6-A8) the joint probability for values $\lambda_1$, $\lambda_2$ 
in the peculiar points which can be knots or saddles (not focuses) is
\[
P(\lambda_1,\lambda_2)d\lambda_1d\lambda_2=\frac{8}{(4-\sqrt{2})\sqrt{\pi}}
|\lambda_1\lambda_2|(\lambda_1-\lambda_2)
e^{-\lambda_1^2-\lambda_2^2}d\lambda_1d\lambda_2,\hspace{0.2cm}
\lambda_1>\lambda_2.\hspace{0.5cm}(A9)
\]

\vspace{0.5cm}

\begin{center}
{\large\bf Appendix B}
\end{center}

\noindent
{\bf Genus curve}

In this appendix we obtain the differential and integrated Genus curve for 
the two-dimensional random Rayleigh field.

According to section 2.2 the value $P=\sqrt{q^2+u^2}$ is the non-linear 
combination of two different independent random Gaussian fields $q$ and $u$. 
Equation (17) for the joint probability distribution of the values $q$, $u$, 
$q_i$, $u_i$ $q_{ij}$, $u_{ij}$, $i,j=1,2$ contains the quadratic form $A$ 
and $\det M$, where $M$ is a covariance matrix:
\[
\begin{array}{l}
A=q^2+u^2+2(q_1^2+q_2^2+u_1^2+u_2^2)+
\frac{(q_{11}+q_{22}+\gamma q)^2+(u_{11}+u_{22}+\gamma u)^2}{1-\gamma^2}+\\
+2(q_{11}-q_{22})^2+2(u_{11}-u_{22})^2+8q^2_{12}+8u^2_{12},\\
\det M=2^{-12}(1-\gamma^2)^2.\\
\end{array}
\hspace{1.2cm}(B1)
\]
The substitutions in Eq.(17)
\[
\begin{array}{ll}
q=p\cos\varphi, & u=p\sin\varphi,\\
p_i=q_{i}\cos\varphi+u_{i}\sin\varphi, & l_i=q_{i}\sin\varphi-u_{i}
\cos\varphi,\\
l_{ij}=q_{ij}\cos\varphi+u_{ij}\sin\varphi, & \tilde{l}_{ij}=q_{ij}
\sin\varphi-u_{ij}\cos\varphi,\\
i,j=1,2, & \\
\end{array}
\hspace{3.8cm}(B2)
\]
and integration over $dl_id\tilde{l}_{ij}$
  gives us the joint probability for the 
distribution of the values $p$, $p_i$, $l_i$, $l_{ij}$:
\[
X(p, p_i, l_i, l_{ij})dpdp_idl_idl_{ij}=
\frac{32}{(2\pi)^{7/2}(1-\gamma^2)^{1/2}}pe^{-\frac{\tilde{A}}{2}}dpdp_idl_i
dl_{ij}\hspace{0.1cm}i,j=1,2
\hspace{0.3cm}(B3)
\]
Following Eqs.(16) and (B2) we can get the expression:
\[
pp_{ij}=pl_{ij}+\gamma l_il_j;\hspace{10.cm}(B4)
\]
using Eqs.(20), (B3) and (B4) we obtain
\[
g(p)=\frac{32}{(2\pi)^{7/2}(1-\gamma^2)^{1/2}}\int(p_{11}p_{22}-p_{12}^2)
pe^{-\frac{\tilde{\tilde{A}}}{2}} dl_idp_{ij}\hspace{0.5cm} i,j=1,2
\]
\[
\tilde{\tilde{A}}=p^2+2(l_1^2+l_2^2)+\frac{(p_{11}+p_{22}-
\gamma a)^2}{1-\gamma^2}+
2(p_{11}-p_{22}-\gamma b)^2+8(p_{12}-\gamma c)^2 \hspace{1cm}(B5)
\]
\[
a=\frac{l_1^2+l_2^2}{p}-p,\hspace{0.5cm} b=\frac{l_1^2-l_2^2}{p}
\hspace{0.5cm}   c=\frac{l_1^2l_2^2}{p}
\]
the integration over $dp_{ij}$, $dl_i$ gives us differential Genus curve
\[
g(p)=\frac{1}{4\pi}\left(\frac{\sigma_1}{\sigma_0}\right)^2p(p^2-3)
e^{-\frac{p^2}{2}}\hspace{8cm}(B6)
\]
The integrated curve is 
\[
G(p)=\int\limits^{\infty}_{p}g(p')dp'=\frac{1}{4\pi}\left(\frac{\sigma_1}
{\sigma_0}\right)^2
(p^2-1)e^{-\frac{p^2}{2}}.\hspace{5.3cm}(B7)
\]

\begin{center}
{\bf REFERENCES}\\
\end{center}
Adler, R.J., The geometry of random fields, John Wiley \& Sons, 
Chichester, 1981.\\
Arbuzov P., Kotok E., Naselsky P. and Novikov I., 1997a, Preprint TAC, 
1997-017, Intern. J.of Mod.Physics (submitted)\\
Arbuzov P., Kotok E., Naselsky P. and Novikov I., 1997b, Preprint TAC, 
1997-021, Intern. J.of Mod.Physics (submitted).\\
Bardeen,J.M.,Bond,J.R.,Kaiser,N.,\& Szalay, A.S., Ap.J. {\bf 304},(1986)
15-61.\\
Basco M.A., Polnarev A.G., Sov. Astron., 1979, {\bf 24}, 3\\
Bennett C.L., et al, 1996,  ApJ. {\bf 464}, L1.\\
Bond J.R., G. Efstathiou., 1987, M.N.R.A.S. {\bf 336}, 655\\
Cheng E.S., et al, 1994, ApJ. {\bf 422}, L37\\
Coles,P.\& Barrow,J.D., 1987, M.N.R.A.S {\bf 228}, 407-426\\
Coles,P. M.N.R.A.S, 1988, {\bf 231}, 125-130\\
Coulson P., Grittenden. R., Turok N., 1994, Phys. Rev. Lett., {\bf 73}, 2390\\
Dominik, K., \& Shandarin, S. 1992, ApJ, {\bf 393}, 450\\
Doroshkevich,A.G.,Astrophysics {\bf 6} (1970), 320-330\\
De Bernardis. P., et al, 1994, ApJ. {\bf 422}, L33\\
Grittenden R. Coulson P., Turok N., 1995, Phys. Rev. D, {\bf 52}, 5402\\
Hadwiger,H., Vorlesungen uber Inhalt, Oberflache und Isoperimetrie,
Springer Verlag, Berlin, 1957\\
Harari,D.D., \& Zaldarriaga, M. 1993, Phys. Letters B, {\bf 319}, 96\\
Harari,D.D., Hayward,J.D., \& Zaldarriaga, M. 1996, Phys. Rev. D, {\bf 55},
1841\\
Hancock S., et al, 1994, Nature {\bf 367}, 333.\\
Hu. W and M.White, 1997, astro-ph 970647\\
Jungman G., Kamionkowski M.A.  Kosowsky A., D. Spergel, 1996, 
Phys. Rev. D, {\bf 54}, 1332\\
Kamionkowski M.A. Kosowsky A. 
and Stebbins A., 1997, Phys. Rev. D {\bf 55}, 7368\\
Keating, B., Polnarev, A., Steinberger, J., Timbie, P., (1987) astro-ph\\
Kogut,A. et al. 1994, ApJ, {\bf 433}, 435\\
Kosowsky A., 1996, Annals Phys. {\bf 246}, 49\\
Luo, X. \& Schramm, D.N. 1994, Phys. Rev. Lett., {\bf 71}, 1124\\
Masi S. et al, 1996, ApJ. {\bf 463}, L47\\
Melott, A.L. 1990, Phys. Reports, {\bf 193}, 1\\
Minkowski,H., Mathematische Annalen {\bf 57} (1903), 447-495\\
Naselsky P. and Novikov D., 1995, ApJ. {\bf 444}, L1\\
Naselsky P.D. and Polnarev A.G., 1987, Astrophysica {\bf 26}, 543\\
Netterfield  et al., 1996, astro-ph 9601197\\
Novikov D. and H. J\o rgensen, 1996a, ApJ. {\bf 471}, 521\\
Novikov D. and H. J\o rgensen, 1996b, Intern. J.of Mod.Physics {\bf 5}, 319\\
Ng K.L. and Ng K.W., 1995, Phys. Rev. D, {\bf 51}, 364\\
Ng K.L. and Ng K.W., 1996, ApJ. {\bf 456}, L1\\
Polnarev A.G., 1985, Sov. Astron., 1979, {\bf 62}, 1041\\
Rees M., 1968,  ApJ. {\bf 153}, L1\\
Scott,D., Silk,J., \& White,W. 1995, Science, {\bf 268}, 829\\
Scott P.F., et al, 1996, ApJ. {\bf 461}, L1\\
Schmalzing J., Gorski K.M., astro-ph/9710185\\
Seljak, U., \& Zaldarriaga, M. Ap.J., (1996), {\bf 469}, 437\\
Seljak, U., \& Zaldarriaga, M. Ap.J., (1996), astro-ph 9609169\\
Shandarin, S.F. 1983, Soviet Astron. Lett., {\bf 9}, 104\\
Smoot G., et al., 1992,   ApJ. Lett. {\bf 396}, L1\\
Smoot G., et al., 1994,   ApJ. {\bf 437}, 1\\
Tanaka et al. 1995, astro-ph 9512067\\
Torres, S., et al., 1995, MNRAS, {\bf 274}, 853-857\\
Winitzki, S., \& Kosowsky, A. (1997) astro-ph/9710164\\
Zaldarriaga, M., Harari, D., 1995, Phys. Rev. D, {\bf 52}, 3276\\
Zaldarriaga, M., \& Seljak, U., 1997, Phys. Rev. D, {\bf 55}, 1830\\
Zaldarriaga, M. 1997, Phys. Rev. D, {\bf 55}, 1822\\
Zeldovich, Ya.B. 1982, Soviet Astron. Lett., {\bf 8},102.

\clearpage


\begin{figure}
\begin{center}
\vspace{0.3cm}\hspace{2cm}\epsfxsize=11cm 
\epsfbox{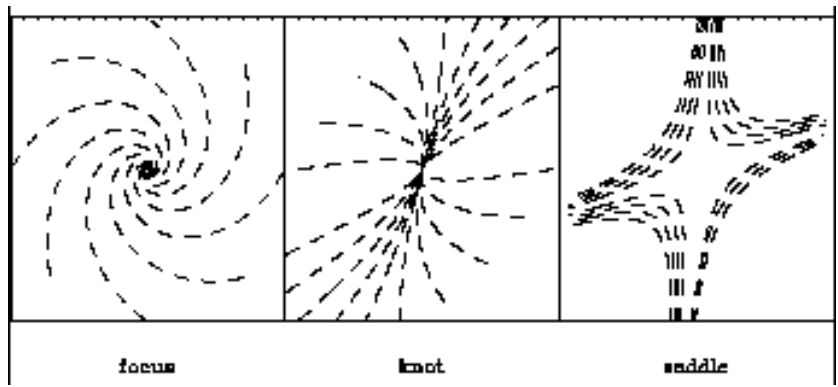}
\end{center}
    \caption{Classification of singular points. 
Field of polarization around singular points $p=0$.
Dashed lines 
show the direction of the pseudo-vector $\overline{P}$ (but not its value).
Left - focus, middle - knot, right - saddle.}
\end{figure}



\begin{figure}
\begin{center}
\leavevmode\psfig{figure=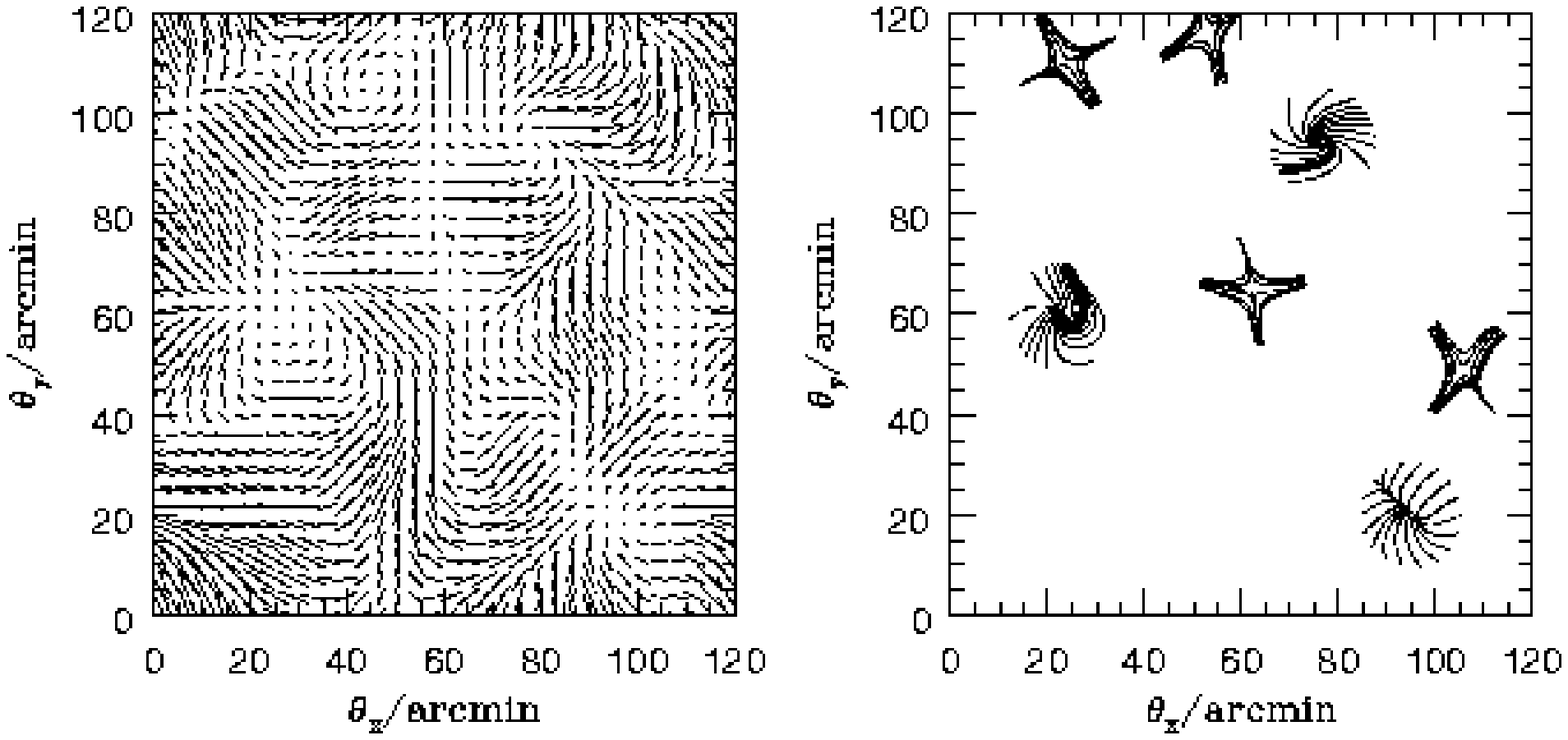		,height=16cm}
\end{center}
	\caption{
Simulated map $2^0\times 2^0$ of the CMB polarization field
for the scale-invariant adiabatic CDM model with $\Omega=1$, 
$\Omega_b=0.03$, h=0.75 with smoothing angle 5 arcmin (FWHM). The
simulation technique for small parts of the sky and spectrum for 
simulations are from Bond and Efstaphiou 1987.
\hspace{1cm}Left - polarization field.
The length of each vector is proportional to the degree of polarization
and the orientation gives the plane of polarization. For visual clarity,
we only use $50 \times 50$ vectors.
\hspace{1cm}Right - the same as left, 
but we plot only the orientation of polarization
in the vicinity of non-polarized points (solid lines). This map containes
7 non-polarized points - 2 foci, 1 knot and 4 saddles.
}
\end{figure}



\begin{figure}
\begin{center}
  \leavevmode\psfig{figure=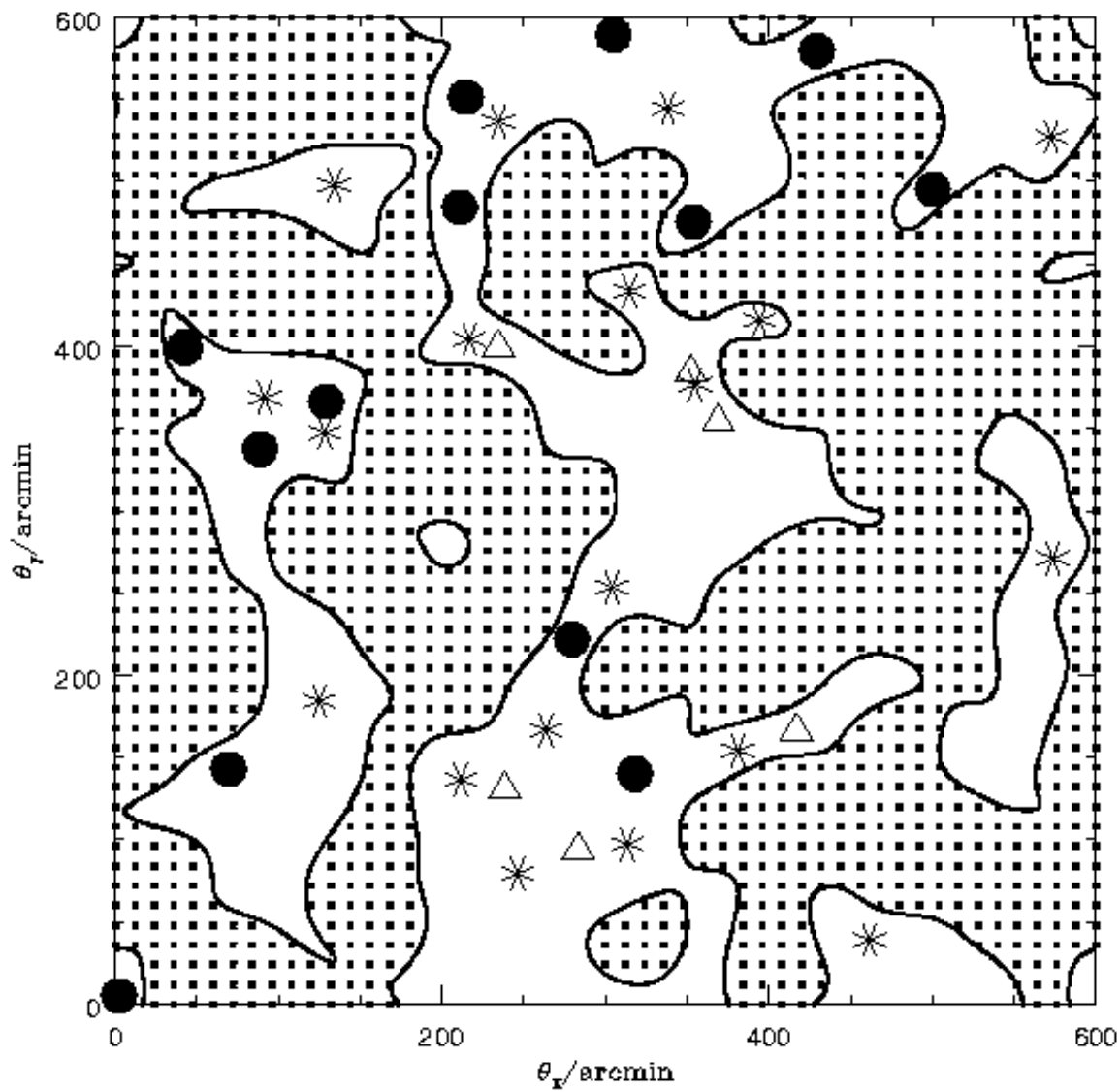	,height=16cm}
\end{center}
\caption{
Simulated map $10^0 \times 10^0$ of CMB polarization for
the same model as in the
fig. 2. Dashed area corresponds to the regions with polarization
degree $p>p_t$. Solid lines are the boundary between regions with
$p>p_t$ and  $p<p_t$. Circles, threangles and stars are foci, knots and
saddles correspondingly. This map containes 13 fosi, 6 knots and 19 
saddles.}
\end{figure}



\begin{figure}
\begin{center}
  \leavevmode\psfig{figure=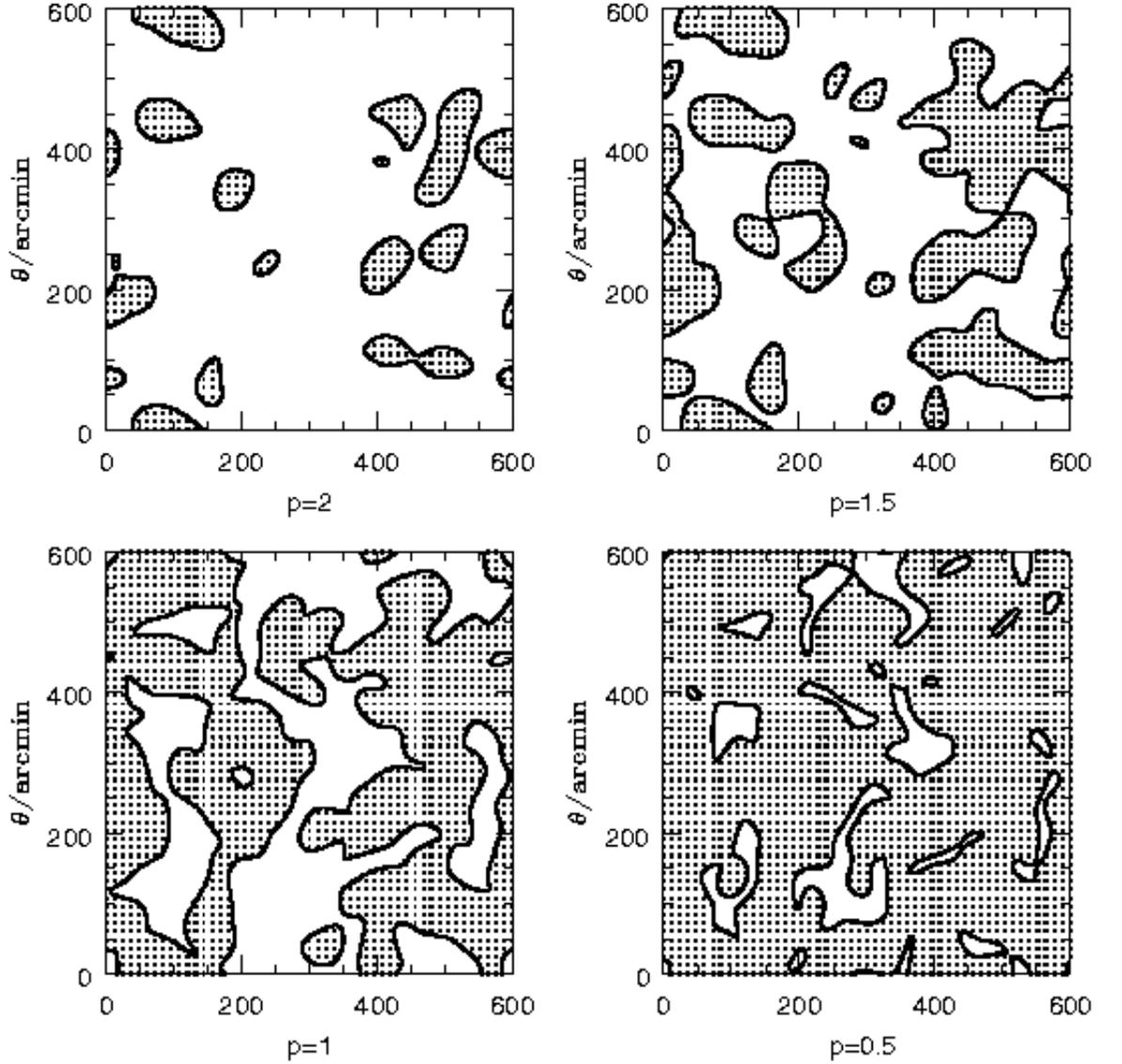	,height=16cm}
\end{center}
\caption{
The same as in fig. 3, but without non-polarized points. We  
plot area, where $p>p_t$ for different values of $p_t$: $p_t=2, 1.5, 1, 0.5$.
Spots with $p>p_t$ percolate then $p_t=1$, which corresponds to the
$e^{-1/2} \sim 61\%$ of the maps area.}
\end{figure}


\begin{figure}
\begin{center}
  \leavevmode\psfig{figure=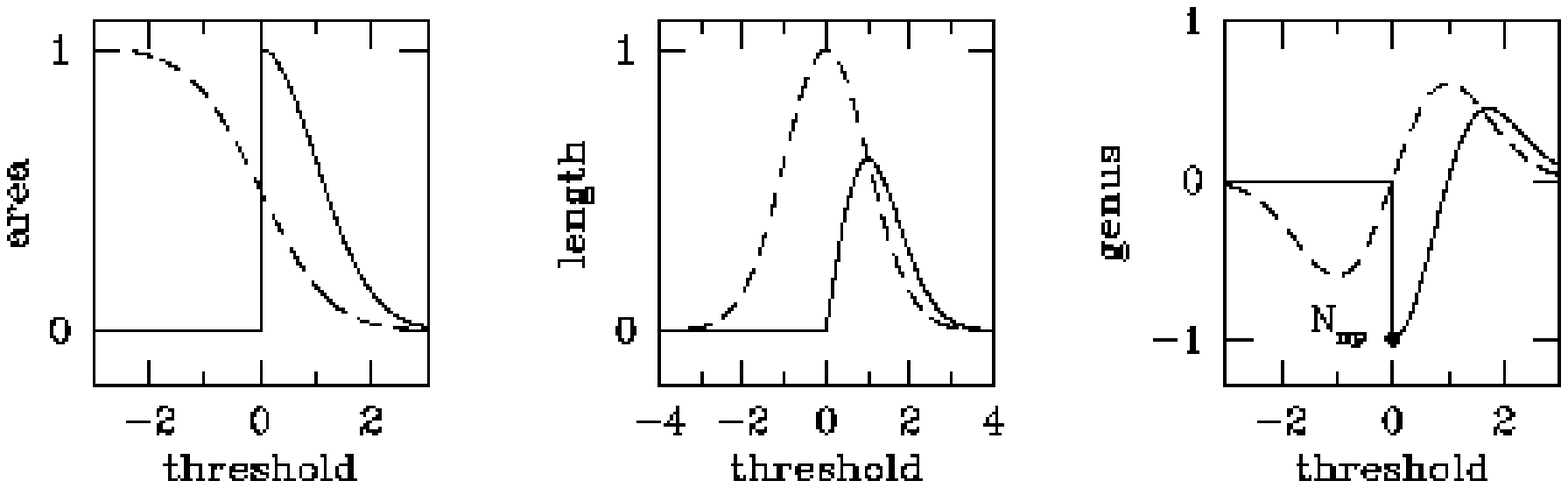	,height=16cm}
\end{center}
	\caption{
Minkowski functionals for CMB polarization (solid lines) and anisotropy
(dashed lines). Threeshold is given in the units of $\sigma_0$
for polarization and in the units of 
$\sqrt{\langle (\Delta T)^2\rangle}$ for anisotropy.}
\end{figure}


\end{document}